\documentclass[aps,preprint,showpacs]{revtex4-1}
\usepackage{amsmath,graphicx,bbm,mathrsfs,amssymb,pst-all,bm,color}

\begin{document}

\title{Efficient quantum algorithms for solving quantum linear system problem%
}
\author{Hefeng Wang$^1$}
\email{wanghf@mail.xjtu.edu.cn}
\author{Hua Xiang$^2$}
\email{hxiang@whu.edu.cn}
\affiliation{$^{1}$Department of Applied Physics, School of Physics, Xi'an
Jiaotong University and Shaanxi Province Key Laboratory of Quantum
Information and Quantum Optoelectronic Devices, Xi'an, 710049, China\\
MOE Key Laboratory for Nonequilibrium Synthesis and Modulation of Condensed Matter, Xi'an Jiaotong University, Xi'an, 710049, China}
\affiliation{$^{2}$School of Mathematics and Statistics, Wuhan University,
Wuhan, 430072, China}

\begin{abstract}
We transform the problem of solving linear system of equations $A\mathbf{x}=%
\mathbf{b}$ to a problem of finding the right singular vector with singular
value zero of an augmented matrix $C$, and present two quantum algorithms
for solving this problem. The first algorithm solves the problem directly by
applying the quantum eigenstate filtering algorithm with query complexity of
$O\left( s\kappa \log \left( 1/\epsilon \right) \right) $ for a $s$-sparse
matrix $C$, where $\kappa $ is the condition number of the matrix $A$, and $%
\epsilon $ is the desired precision. The second algorithm uses the quantum
resonant transition approach, the query complexity scales as $O\left[
s\kappa + \log\left( 1/\epsilon \right)/\log \log \left( 1/\epsilon \right) %
\right] $. Both algorithms meet the optimal query complexity in $\kappa $,
and are simpler than previous algorithms.
\end{abstract}

\maketitle

\section{Introduction}

Solving linear system of equations~(LSE) is one of the fundamental problems
in scientific computation. Given an $N\times N$ matrix $A$ and a vector $%
\mathbf{b}$, the task is to find a vector $\mathbf{x}$ such that $A\mathbf{x}%
=\mathbf{b}$. Solving high-dimensional LSE is expensive on a classical
computer. Classical linear solvers can be categorized into the direct
methods and the iterative methods~\cite{Golub0, demmel}. The direct methods
such as Gaussian elimination solve LSE with runtime scales as $O(N^{3})$.
There exists more efficient classical linear system solver that scales as $%
O(N^{\nu })$ where $\nu \leqslant 2.373$~\cite{CW, Gall}, but it is
difficult to utilize in practice due to numerical instability. The iterative
methods show great advantages when they converge quickly, the iteration
number is an indicator for the efficiency of these methods. E.g., for a
symmetric positive-definite problem, the steepest descent method needs $%
O(\kappa \log (1/\epsilon ))$ iterations and the conjugate gradient method
needs $O(\sqrt{\kappa }\log (1/\epsilon ))$, where $\kappa $ is the
condition number of the matrix $A$ defined as the ratio between the largest
and the smallest singular values of $A$, or $\left\Vert A\right\Vert
\left\Vert A^{-1}\right\Vert $, where $\left\Vert \cdot \right\Vert $
denotes vector or matrix $2$-norm, and $\epsilon $ is the desired precision
of the solution.

Quantum computation provides an efficient way of solving quantum linear
system problem~(QLSP), which aims to prepare a quantum state that is
proportional to the solution vector of a given LSE. Quantum algorithms for
solving the QLSP either apply the matrix inversion operator $A^{-1}$
directly on the state $|b\rangle $ to obtain $|x\rangle $, e.g. the HHL
algorithm~\cite{HHL, childs}, or transform the LSE to an eigenvalue problem
where the ground state of the problem Hamiltonian is the state $|x\rangle $~%
\cite{AQCLSE, Lin}. Recently, a quantum eigenstate filtering~(QEF) algorithm~%
\cite{Lin} was proposed for solving the QLSP by combining with adiabatic
quantum computing~(AQC) or quantum Zeno effect~(QZE), and achieves near
optimal query complexity of $O\left( s\kappa \text{poly}\log \left( \kappa
\right) \log \left( 1/\epsilon \right) \right) $, where $s$ is the sparsity
of the matrix $A$. The QEF algorithm approximates a spectral projection
operator by using the quantum signal processing~(QSP)~\cite{QSP} method and
project out the quantum state $|x\rangle $ from an initial state prepared
through an AQC procedure. Such a procedure can also be realized through QZE
by applying a sequence of QEF procedures following the same path as that of
the AQC method.

In this work, we solve the QLSP in a simpler way. We first prove that the
solution vector to the LSE is contained in a vector proportional to the
right singular vector of the augmented matrix $C=\left(
\begin{array}{cc}
A & \beta ^{-1}\mathbf{b}%
\end{array}%
\right) $ with corresponding singular value $0$, where $\beta $ is a
parameter. Then we construct a Hermitian matrix
\begin{equation}
B=\left(
\begin{array}{cc}
0 & C \\
C^{\dag } & 0%
\end{array}%
\right) ,
\end{equation}%
and transform the QLSP to a problem of finding the eigenstate of the matrix $%
B$ with eigenvalue $0$. In our previous work, we proposed a quantum
algorithm for finding an eigenstate with known eigenvalue of a Hermitian
matrix through quantum resonant transition~(QRT)~\cite{whf0, whf1}. Here, we
set a simple initial state whose overlap with the desired eigenstate of the
matrix $B$ is $O\left( 1\right) $, then both the QEF and the QRT algorithms
can be applied for solving the QLSP. The QEF algorithm has query complexity
of $O\left( s\kappa \log \left( 1/\epsilon \right) \right) $, and the query
complexity of the QRT algorithm scales as $O\left[ s\kappa +\log \left(
1/\epsilon \right) /\log \log \left( 1/\epsilon \right) \right] $ for
solving the problem. Both algorithms achieve the optimal query complexity in
$\kappa $. In previous quantum algorithms, the complexity for preparing the
state $|b\rangle $ and querying the matrix $A$ are considered separately,
while they are combined together in our work. The algorithms do not need to
use complex procedures such as phase estimation, amplitude amplification,
AQC or QZE, thus they are easier to implement experimentally than previous
algorithms.

\section{LSE and singular value decomposition}

The LSE $A\mathbf{x}=\mathbf{b}$ can be written in form of
\begin{equation}
C\left(
\begin{array}{c}
\mathbf{x} \\
-\beta%
\end{array}%
\right) =0.
\end{equation}%
By performing singular value decomposition~(SVD), we have $C=SDV^{\dag }$,
where $S$ and $V$ are unitary matrices of dimension $N$ and $(N+1)$,
respectively, and $D=\left(
\begin{array}{cc}
D^{\prime } & \mathbf{0}%
\end{array}%
\right) $, $D^{\prime }=$diag$(\sigma _{1},\ldots ,\sigma _{N})$, and $%
\mathbf{0}$ is the zero column vector of dimension $N$, $\sigma
_{1}\geqslant \cdots \geqslant \sigma _{N}$\ are the singular values of the
matrix $C$. Correspondingly, $\mathbf{s}_{1},\ldots ,\mathbf{s}_{N}$ are
column vectors of $S$, and $\mathbf{v}_{1},\ldots ,\mathbf{v}_{N+1}$ are
column vectors of $V$.

\textbf{Theorem 1}\textit{.--Suppose }$\mathbf{x}$\textit{\ is the solution
to the equation }$A\mathbf{x}=\mathbf{b}$\textit{, where }$A$\textit{\ is an
}$N\times N$\textit{\ nonsingular matrix, }$\mathbf{v}_{N+1}$ \textit{is the
right singular vector corresponding to} $C\mathbf{v}_{N+1}=\mathbf{0}$%
\textit{, where} $C=\left(
\begin{array}{cc}
A & \beta ^{-1}\mathbf{b}%
\end{array}%
\right) $\textit{, then the vector} $\left(
\begin{array}{cc}
\mathbf{x^{\text{\textbf{T}}},} & -\beta%
\end{array}%
\right) ^{\text{\textbf{T}}}$ \textit{that satisfies Eq.}~($1$)\textit{\ is
proportional to the singular vector} $\mathbf{v}_{N+1}$ \textit{of the matrix%
} $C$.

\textit{Proof.--}Since the rank of the matrix $A$ is $N$, let $\sigma
_{N+1}=0$, the singular values of the matrix $C$ satisfy $\sigma
_{1}\geqslant \cdots \geqslant \sigma _{N}>\sigma _{N+1}=0$. The matrix $C$
can be written as $C\mathbf{=}\sum_{i=1}^{N}\sigma _{i}\mathbf{s}_{i}\mathbf{%
v}_{i}^{\dag }$. The column vectors $\mathbf{v}_{1},\ldots ,\mathbf{v}_{N+1}$
of the unitary matrix $V$ are orthogonal to each other, therefore to satisfy
Eq.~($1$), the vector $\left(
\begin{array}{cc}
\mathbf{x^{\text{\textbf{T}}},} & -\beta%
\end{array}%
\right) ^{\text{\textbf{T}}}$ must be orthogonal to the vectors $\mathbf{v}%
_{1},\ldots ,\mathbf{v}_{N}$, thus it is proportional to the vector $\mathbf{%
v}_{N+1}$. $\square $

Let $\left\Vert A\right\Vert \leq 1$, we perform SVD for the matrix $A$, $%
\bar{S}^{\dag }A\bar{V}=$diag$\left( \bar{\sigma}_{1},\bar{\sigma}%
_{2},\ldots ,\bar{\sigma}_{N}\right) $. The singular values of the matrices $%
A$ and $C$ satisfy the following relation~\cite[Corollary 8.6.3, p487]%
{Golub0}:
\begin{equation}
\sigma _{1}\geqslant \bar{\sigma}_{1}\geqslant \sigma _{2}\geqslant \bar{%
\sigma}_{2}\ldots \geqslant \bar{\sigma}_{N}>\sigma _{N+1}=0.
\end{equation}%
Then the energy gap between the singular states $\mathbf{v}_{N+1}$ and $%
\mathbf{v}_{N}$ is
\begin{equation}
\Delta =\sigma _{N}-\sigma _{N+1}\geqslant \Delta ^{\ast }=\bar{\sigma}_{N}=%
\frac{1}{\kappa }.
\end{equation}%
Suppose $\left\Vert \mathbf{b}\right\Vert =1$, then $\left\Vert \mathbf{x}%
\right\Vert =\left\Vert A^{-1}\right\Vert \left\Vert \mathbf{b}\right\Vert $
is in the range of $\left[ 1,\kappa \right] $. The vector $\left(
\begin{array}{cc}
\mathbf{x^{\text{\textbf{T}}},} & -\beta%
\end{array}%
\right) ^{\text{\textbf{T}}}$ can be normalized as $\mathbf{v}%
_{N+1}=d_{0}\left(
\begin{array}{cc}
\mathbf{x^{\text{\textbf{T}}}/\left\Vert \mathbf{x}\right\Vert ,} & 0%
\end{array}%
\right) ^{\text{\textbf{T}}}+d_{1}\left(
\begin{array}{cc}
\mathbf{0^{\text{\textbf{T}}},} & 1%
\end{array}%
\right) ^{\text{\textbf{T}}}$, where $d_{0}/d_{1}=\mathbf{\left\Vert \mathbf{%
x}\right\Vert /}\beta $.

In the algorithm, the solution state $\mathbf{x/\left\Vert \mathbf{x}%
\right\Vert }$ is obtained in two steps, we first obtain the state $\mathbf{v%
}_{N+1}$, then extract the solution state $\mathbf{x/\left\Vert \mathbf{x}%
\right\Vert }$ from $\mathbf{v}_{N+1}$. The choice of the parameter $\beta $
has dual influence on both the overlap of $\mathbf{v}_{N+1}$ with the
initial state and the post-selection of solution state. We first set the
parameter $\beta $ in the order of $O\left( \kappa \right) $, thus the
component $d_{1}$ is guaranteed to be in the order of $O\left( 1\right) $,
and run the algorithm to obtain the state $\mathbf{v}_{N+1}$. We run the
algorithm and perform measurement on the state $\mathbf{v}_{N+1}$ for a
number of times to estimate the value of $d_{1}$. This can be achieved by
measuring the $(N+1)$-th component of the vector $\mathbf{v}_{N+1}$ to
obtain the probability of its outcome being $1$, and obtain $d_{0}=\sqrt{%
1-d_{1}^{2}}$. After that we reset the parameter $\beta $ such that $\beta $
is in the order of $\mathbf{\left\Vert \mathbf{x}\right\Vert }$ and run the
algorithm again. Then both the components $d_{0}$ and $d_{1}$ are in the
order of $O\left( 1\right) $. The state $\mathbf{x^{\text{\textbf{T}}%
}/\left\Vert \mathbf{x}\right\Vert }$ can be extracted efficiently from the
state $\mathbf{v}_{N+1}$.

In the following we describe the algorithms for obtaining the state $\mathbf{%
v}_{N+1}$. The matrix $B$ has eigenvalues and eigenstates as
follows: $B|\varphi _{j}\rangle =E_{j}|\varphi _{j}\rangle $, $j=1,\ldots
,2N+1$, and $E_{1}=-\sigma _{1}$, $\ldots $, $E_{N}=-\sigma _{N}$, $%
E_{N+1}=0 $, $E_{N+2}=\sigma _{N}$, $\ldots $, $E_{2N+1}=\sigma _{1}$, and $%
|\varphi _{1}\rangle =(\mathbf{s}_{1},-\mathbf{v}_{1})$, $\ldots $, $%
|\varphi _{N}\rangle =(\mathbf{s}_{N},-\mathbf{v}_{N})$, $|\varphi
_{N+1}\rangle =|\mathbf{v}_{N+1}\rangle $, $|\varphi _{N+2}\rangle =(\mathbf{%
s}_{N},\mathbf{v}_{N})$, $\ldots $, $|\varphi _{2N+1}\rangle =(\mathbf{s}%
_{1},\mathbf{v}_{1}) $. The problem of solving the LSE is transformed to
finding the eigenstate $|\mathbf{v}_{N+1}\rangle =(\mathbf{0},\mathbf{v}%
_{N+1})$ of the matrix $B$ with eigenvalue $0$. The gap between the
eigenstate $|\mathbf{v}_{N+1}\rangle $ and its nearest neighboring
eigenstate $|\varphi _{N+2}\rangle $ is in the order of $O(\kappa ^{-1})$ in
the worst case. We set the initial state of the quantum circuit as $|\mathbf{%
1}\rangle =(\mathbf{0},\mathbf{0},1)$, whose overlap with the desired
eigenstate $|\mathbf{v}_{N+1}\rangle $ of the matrix $B$ is $d_{1}=\langle
\mathbf{1}|\mathbf{v}_{N+1}\rangle $ which scales as $O(1)$ in the
algorithm. We then apply the QEF and the QRT algorithms for obtaining the
state $|\mathbf{v}_{N+1}\rangle $ from the initial state $|\mathbf{1}\rangle
$, and extract the solution state $\mathbf{x/\left\Vert \mathbf{x}%
\right\Vert }$ from the state $\mathbf{v}_{N+1}$.

\section{Solving QLSP through QEF}

The QEF algorithm projects out the desired eigenstate of a Hermitian matrix
from an initial state by implementing an eigenstate-filtering function using
the QSP method, which is a powerful algorithm for implementing a polynomial
function of matrices on a quantum computer with minimal number of ancilla
qubits. The matrix $B$ is encoded in a unitary matrix by using the
block-encoding technique, then the QSP method is applied to implement the
QEF function to project out the eigenstate $|\mathbf{v}_{N+1}\rangle $ from
the initial state.

The matrix $B$ is represented on an $n$-qubit quantum register, an ($m+n$%
)-qubit unitary operator $U_{B}$ is called a $(\alpha ,m,\epsilon )$%
-block-encoding of the matrix $B$~\cite{Qubitize}, if
\begin{equation}
\left\Vert B-\alpha \left( \langle 0^{m}|\otimes I_{n}\right) U_{B}\left(
|0^{m}\rangle \otimes I_{n}\right) \right\Vert \leq \epsilon ,
\end{equation}%
where $I_{n}$ is an $n$-qubit identity matrix. Block-encoding of $B$ can
also be written in form of
\begin{equation}
U_{B}=\left(
\begin{array}{cc}
B/\alpha & \cdot \\
\cdot & \cdot%
\end{array}%
\right) ,
\end{equation}%
where $\alpha $ is a parameter such that $\left\Vert B/\alpha \right\Vert
\leq 1$. It is determined by the largest eigenvalue of the matrix $B$, or
the largest singular value $\sigma _{1}$ of the matrix $C$. We estimate the
upper bound of $\sigma _{1}$ as follows:
\begin{eqnarray}
\sigma _{\max }^{2}(C) &=&\max_{\mathbf{y}\neq 0}\frac{\left\Vert C\mathbf{y}%
\right\Vert ^{2}}{\left\Vert \mathbf{y}\right\Vert ^{2}}  \notag \\
&=&\max_{\mathbf{y}=\left(
\begin{array}{c}
y_{1} \\
y_{2}%
\end{array}%
\right) \neq 0}\frac{\left\Vert \left(
\begin{array}{cc}
A & \beta ^{-1}\mathbf{b}%
\end{array}%
\right) \left(
\begin{array}{c}
y_{1} \\
y_{2}%
\end{array}%
\right) \right\Vert ^{2}}{\left\Vert \left(
\begin{array}{c}
y_{1} \\
y_{2}%
\end{array}%
\right) \right\Vert ^{2}}  \notag \\
&=&\max_{\left(
\begin{array}{c}
y_{1} \\
y_{2}%
\end{array}%
\right) \neq 0}\frac{\left\Vert Ay_{1}+\beta ^{-1}\mathbf{b}y_{2}\right\Vert
^{2}}{\left\Vert y_{1}\right\Vert ^{2}+\left\Vert y_{2}\right\Vert ^{2}}
\notag \\
&\leq &\max_{\left(
\begin{array}{c}
y_{1} \\
y_{2}%
\end{array}%
\right) \neq 0}\frac{\left( \left\Vert Ay_{1}\right\Vert +\beta
^{-1}\left\Vert \mathbf{b}\right\Vert \left\Vert y_{2}\right\Vert \right)
^{2}}{\left\Vert y_{1}\right\Vert ^{2}+\left\Vert y_{2}\right\Vert ^{2}}.
\end{eqnarray}%
Then we have:
\begin{eqnarray}
\sigma _{\max }\left( C\right) &\leq &\max_{\left(
\begin{array}{c}
y_{1} \\
y_{2}%
\end{array}%
\right) \neq 0}\frac{\left\Vert Ay_{1}\right\Vert +\beta ^{-1}\left\Vert
\mathbf{b}\right\Vert \left\Vert y_{2}\right\Vert }{\sqrt{\left\Vert
y_{1}\right\Vert ^{2}+\left\Vert y_{2}\right\Vert ^{2}}}  \notag \\
&\leq &\max_{\left(
\begin{array}{c}
y_{1} \\
y_{2}%
\end{array}%
\right) \neq 0}\frac{\left\Vert Ay_{1}\right\Vert }{\sqrt{\left\Vert
y_{1}\right\Vert ^{2}+\left\Vert y_{2}\right\Vert ^{2}}}+\max_{\left(
\begin{array}{c}
y_{1} \\
y_{2}%
\end{array}%
\right) \neq 0}\frac{\beta ^{-1}\left\Vert \mathbf{b}\right\Vert \left\Vert
y_{2}\right\Vert }{\sqrt{\left\Vert y_{1}\right\Vert ^{2}+\left\Vert
y_{2}\right\Vert ^{2}}}  \notag \\
&\leq &\max_{y_{1}\neq 0}\frac{\left\Vert Ay_{1}\right\Vert }{\left\Vert
y_{1}\right\Vert }+\max_{y_{2}\neq 0}\frac{\beta ^{-1}\left\Vert \mathbf{b}%
\right\Vert \left\Vert y_{2}\right\Vert }{\left\Vert y_{2}\right\Vert }
\notag \\
&=&\sigma _{\max }\left( A\right) +\beta ^{-1}  \notag \\
&=&\bar{\sigma}_{1}+\beta ^{-1}.
\end{eqnarray}
The parameter $\beta $ is in the range of $\left[ 1,\kappa \right] $, thus
the maximum singular value of the matrix $C$ is no more than $2$.

The polynomial eigenvalue transformation of $B/\alpha $ can be implemented
based on the following theorem~\cite{Lin,QSVT}:

\textbf{Theorem 2} (\textit{Polynomial eigenvalue transformation with
definite parity via quantum signal processing}): \textit{Let }$U_{B}$\textit{%
\ be an }($\alpha $, $m$, $0$)\textit{-block-encoding of the Hermitian
matrix }$B/\alpha $ \textit{and }$P\in
\mathbb{R}
\lbrack w]$ \textit{be a degree-}$l$ \textit{even or odd real polynomial and
}$\left\vert P\left( w\right) \right\vert \leq 1$ \textit{for any} $w\in
\lbrack -1,1]$. \textit{Then there exists a }($1,m+1,0$)\textit{%
-block-encoding} $U_{\vec{\varphi}}$ \textit{of }$P(B/\alpha )$\textit{\
using }$l$\textit{\ queries of }$U_{B}$\textit{, }$U{}_{B}^{\dag }$\textit{,
and }$O((m+1)l)$\textit{\ other primitive quantum gates.}

Let $B/\alpha =\sum_{\lambda }\lambda \left\vert \lambda \right\rangle
\left\langle \lambda \right\vert $, and define $\Pi :=\left\vert
0\right\rangle \left\langle 0\right\vert $ acting on an auxiliary qubit, and
$\Pi _{\phi }:=e^{i\phi \left( 2\Pi -I\right) }$. For $l$ is even, $U_{\vec{%
\varphi}}$ is in form of~\cite{grandunif}
\begin{equation}
U_{\vec{\varphi}}=\prod_{k=1}^{l/2}\left( \Pi _{\varphi
_{2k-1}}U_{B}^{\dagger }\Pi _{\varphi _{2k}}U_{B}\right) =\left(
\begin{array}{cc}
P\left( B/\alpha \right) & \cdot \\
\cdot & \cdot%
\end{array}%
\right) ,
\end{equation}%
where $P\left( B/\alpha \right) =\sum_{\lambda }P\left( \lambda \right)
\left\vert \lambda \right\rangle \left\langle \lambda \right\vert $ is a
polynomial transform of the eigenvalues of $B/\alpha $. The phase factors $%
(\varphi _{1},\ldots ,\varphi _{l})$ can be calculated efficiently~\cite%
{phase1, phase2, phase3, phase4, phase5}. The eigenvalue transform can be
used to project out the desired eigenstate with known eigenvalue and filter
out other unrelated states. For a Hermitian matrix with eigenvalue $\lambda $
that is known to be separated from other eigenvalues by a gap $\Delta >0$,
it has been shown that the following degree-($l=2k$) polynomial
\begin{equation}
R_{k}\left( w;\Delta \right) =\frac{T_{k}\left( -1+2\frac{w^{2}-\Delta ^{2}}{%
1-\Delta ^{2}}\right) }{T_{k}\left( -1+2\frac{-\Delta ^{2}}{1-\Delta ^{2}}%
\right) }
\end{equation}%
is an optimal polynomial for filtering out the unwanted eigenstates~\cite%
{Lin}, where $T_{k}\left( w\right) $ is the $k$th Chebyshev polynomial of
the first kind. By using this polynomial in eigenvalue transform, the system
can be projected onto the eigenspace corresponding to the eigenvalue $%
\lambda $.

The eigenstate $|\mathbf{v}_{N+1}\rangle $ of the matrix $B$ is separated
from the nearest eigenstate of $B$ by a minimum gap of $\Delta ^{\ast
}=\kappa ^{-1}$. We set the initial state of the quantum circuit as $%
\left\vert \mathbf{1}\right\rangle $, and apply the QEF algorithm to project
out the eigenstate $|\mathbf{v}_{N+1}\rangle $, while filtering out all
other eigenstates. The overlap between the states $\left\vert \mathbf{1}%
\right\rangle $ and $|\mathbf{v}_{N+1}\rangle $ is $d_{1}=O\left( 1\right) $%
. Therefore the eigenstate $|\mathbf{v}_{N+1}\rangle $ can be obtained with
success probability $O\left( 1\right) $. The number of qubits required is $%
\left( n+m+1\right) $.

The implementation of the algorithm requires querying the matrix $B$ that
contains the matrix $C$ which is composed of the matrix $A$ and the column
vector $|b\rangle $. The matrix $A$ can be accessed by oracles $O_{A,1}$ and
$O_{A,2}$ as $O_{A,1}\left\vert j,l\right\rangle =\left\vert j,\nu \left(
j,l\right) \right\rangle $, $O_{A,2}\left\vert j,k,z\right\rangle
=\left\vert j,k,A_{jk}\oplus z\right\rangle $, ($j,k,l,z\in \lbrack N]$),
where the oracle $O_{A,1}$ accepts a row $j$ index and calculates the column
index $\nu \left( j,l\right) $ of the $l$th nonzero element in the $j$th row
of the matrix $A$, and the oracle $O_{A,2}$ accepts a row $j$ and a column $%
k $ index and returns the matrix element $A_{jk}$~\cite{bc}. The vector $b$
is prepared by using an oracle $O_{b}$ as $O_{b}\left\vert 0\right\rangle
=\left\vert b\right\rangle $, and its elements can be accessed by an oracle $%
O_{b,1}$ as $O_{b,1}\left\vert k,z\right\rangle =\left\vert k,b_{k}\oplus
z\right\rangle $, ($k,z\in \lbrack N]$), where the oracle $O_{b,1}$ accepts
the input $k$ and returns the $k$th element $b_{k}$ of the vector $%
\left\vert b\right\rangle $. The matrix $C$ contains the matrix $A$ and the
vector $\left\vert b\right\rangle $ as the $\left( N+1\right) $-th column
vector. If the matrix $C$ is $s$-sparse, then it can be accessed by using
oracles $O_{C,1}$ and $O_{C,2}$ similar to the oracles for accessing the
matrix $A$ above as $O_{C,1}\left\vert j,l\right\rangle =\left\vert
j,f\left( j,l\right) \right\rangle $, $O_{C,2}\left\vert j,k,z\right\rangle
=\left\vert j,k,C_{jk}\oplus z\right\rangle $, ($k,l,z\in \lbrack N],j\in
\lbrack N+1]$), where the oracle $O_{C,1}$ calculates $f\left( j,l\right) $
which is the column index of the $l$th nonzero element in the $j$th row of
the matrix $C$, and the oracle $O_{C,2}$ accepts the input $\left(
j,k\right) $ and returns the matrix element $C_{jk}$ of the matrix $C$. In
this case, a ($s$, $n+2$, $0$)-block-encoding of $C$~can be constructed by
using the $O_{C,1}$ and $O_{C,2}$~\cite{childs, berry}. The complexity of
the QEF algorithm scales as $O\left( \left( \alpha /\Delta ^{\ast }\right)
\log \left( 1/\epsilon \right) \right) $~\cite{Lin}, therefore the
complexity of using the QEF algorithm for obtaining the state $|\mathbf{v}%
_{N+1}\rangle $ scales as $O\left( s\kappa \log \left( 1/\epsilon \right)
\right) $ by querying the oracles $O_{C,1}$ and $O_{C,2}$.

\section{Solving QLSP via QRT}

In Ref.~\cite{whf0, whf1}, we proposed a quantum algorithm for finding an
eigenstate with known corresponding eigenvalue of a Hamiltonian based on
quantum resonant transition. By coupling a probe qubit to a system, a
resonant transition occurs when the transition frequency of the probe qubit
matches a transition in the system, and the system is evolved to the
eigenstate with known eigenvalue. This algorithm can be applied for solving
the QLSP by obtaining the eigenstate $|\mathbf{v}_{N+1}\rangle $ of the
matrix $B$. The algorithm requires $n+1$ qubits with one probe qubit and an $%
n$-qubit register $R$ representing the matrix $B$. The Hamiltonian of the
algorithm is constructed as%
\begin{equation}
H=-\frac{1}{2}\omega \sigma _{z}\otimes I_{n}+H_{R}+c\sigma _{x}\otimes
I_{n},
\end{equation}%
where
\begin{equation}
H_{R}=\varepsilon _{0}|1\rangle \langle 1|\otimes |\mathbf{1}\rangle \langle
\mathbf{1}|+|0\rangle \langle 0|\otimes B\mathbf{,}
\end{equation}%
and $\sigma _{x}$ and $\sigma _{z}$ are the Pauli matrices. The first term
in Eq.~($11$) is the Hamiltonian of the probe qubit, the second term
contains the Hamiltonian of the register $R$ and describes the interaction
between the probe qubit and $R$, and the third term is a perturbation. The
parameter $\varepsilon _{0}$ is used as a reference point to the eigenstate $%
|\mathbf{v}_{N+1}\rangle $ of the matrix $B$ with eigenvalue $0$, and $c\ll 1
$. The condition for resonant transition between states $|1\rangle |\mathbf{1%
}\rangle $ and $|0\rangle |\mathbf{v}_{N+1}\rangle $ is satisfied as $%
E_{N+1}-\varepsilon _{0}=\omega $, that is, $\omega =-\varepsilon _{0}$.

The algorithm is run as follows:

$i$) Set the initial state of the $n+1$ qubits as $|1\rangle |\mathbf{1}%
\rangle $, which is an eigenstate of $H_{R}$ with eigenvalue $\varepsilon
_{0}$.

$ii$) Implement the unitary operator $U(t)=\exp \left( -iHt\right) $ by
setting $\omega =1$ and $\varepsilon _{0}=-1$.

$iii$) Read out the state of the probe qubit.

As the resonant transition occurs, the system is approximately in state $%
\sqrt{1-p}|1\rangle |\mathbf{1}\rangle +\sqrt{p}|0\rangle |\mathbf{v}%
_{N+1}\rangle $, where $p=\sin ^{2}\left( ctd_{1}\right) $ is the decay
probability of the probe qubit, and $c<\Delta ^{\ast }$ and $d_{1}=\langle
\mathbf{1}|\mathbf{v}_{N+1}\rangle $. By performing a measurement on the
probe qubit, if the probe decays to its ground state $|0\rangle $, it
indicates that a resonant transition occurs and the system evolves to the
state $|0\rangle |\mathbf{v}_{N+1}\rangle $; otherwise if the probe qubit
stays in state $|1\rangle $, it means that the register $R$ remains in state
$|\mathbf{1}\rangle $, then we repeat steps $ii$)-$iii$) until the probe
qubit decays to its ground state $|0\rangle $. The number of times the
procedures need to be repeated is proportional to $1/p$.

Errors are introduced by excitations from the initial state $|1\rangle |%
\mathbf{1}\rangle $ to the states $|0\rangle |\varphi _{j}\rangle $ ($%
j=N+2,\cdots ,2N+1$). Similar to the method in Ref.~\cite{wyx}, we estimate
errors introduced when running the algorithm with the evolution time $t=\pi
/(2cd_{1})$ by setting $\varepsilon _{0}=-1$ and $\omega =1$ as follows. Let
\begin{equation}
H_{0}=-\frac{1}{2}\sigma _{z}\otimes I_{n}-|1\rangle \langle 1|\otimes |%
\mathbf{1}\rangle \langle \mathbf{1}|+|0\rangle \langle 0|\otimes B,
\end{equation}%
then the algorithm Hamiltonian $H$ can be written as $H=H_{0}+c\sigma
_{x}\otimes I_{n}$. The Hamiltonian $H_{0}$ is the unperturbed term and has
eigenstates%
\begin{equation}
H_{0}|1\rangle |\mathbf{1}\rangle =\!-\frac{1}{2}|1\rangle |\mathbf{1}%
\rangle ,
\end{equation}%
and%
\begin{equation}
H_{0}|0\rangle |\varphi _{j}\rangle =\left( -\frac{1}{2}\!+\!E_{j}\right)
|0\rangle |\varphi _{j}\rangle ,
\end{equation}%
where $j=1$, $\ldots $, $2N+1$. The system is initialized in state $%
|1\rangle |\mathbf{1}\rangle $.

For a probe qubit coupled to a two-level system described by the Hamiltonian
in Eq.~($11$), the maximum transition probability from the ground state to
the excited state of the two-level system becomes higher as the transition
frequency between the two-level system gets closer to the frequency of the
probe qubit. Therefore the upper bound of the transition probability from
the initial state to the states other than $|\varphi _{N+1}\rangle $ can be
estimated by assuming all the other states are degenerate at the eigenstate $%
|\varphi _{N+2}\rangle $ and without considering competition of the
transition from the initial state to the target state $|0\rangle |\varphi
_{N+1}\rangle $. Then the upper bound of the error of the algorithm is the
transition probability from the initial state to the state $|0\rangle
|\varphi _{N+2}\rangle $ described by the following Hamiltonian in basis of $%
\left\{ |1\rangle |\mathbf{1}\rangle ,|0\rangle |\varphi _{N+2}\rangle
\right\} $:%
\begin{equation}
{H_{err}=}\left(
\begin{array}{cc}
-\frac{1}{2} & c\sqrt{1-d_{1}^{2}} \\
c\sqrt{1-{d}_{1}^{2}} & -\frac{1}{2}+E_{N+2}%
\end{array}%
\right) .
\end{equation}%
When the coefficient $c$ is much less than the energy difference between the
ground state and the first excited state of $H_{0}$, the excitation to the
excited states can be described by the Rabi's formula~\cite[p414]{cohen}.
The transition probability from the initial state $|1\rangle |\mathbf{1}%
\rangle $ to the state $|0\rangle |\varphi _{N+2}\rangle $ is
\begin{equation}
\frac{4c^{2}\left( 1-d_{1}^{2}\right) }{4c^{2}\left( 1-d_{1}^{2}\right)
+E_{N+2}^{2}}\sin ^{2}\left[ \frac{t}{2}\sqrt{4c^{2}\left(
1-d_{1}^{2}\right) +E_{N+2}^{2}}\right] .
\end{equation}%
The upper bound of the transition probability from the initial state to the
states $|0\rangle |\varphi _{j}\rangle $ ($j=N+2,\cdots ,2N+1$) can be
estimated as
\begin{equation}
\sum_{j=N+2}^{2N+1}p_{j}\leq \frac{4c^{2}\left( 1-d_{1}^{2}\right) }{%
4c^{2}\left( 1-d_{1}^{2}\right) +E_{N+2}^{2}}<\frac{4c^{2}\left(
1-d_{1}^{2}\right) }{E_{N+2}^{2}}=\frac{4c^{2}\left( 1-d_{1}^{2}\right) }{%
\sigma _{N}^{2}},
\end{equation}%
which can be controlled as a small number by setting the coefficient $c$ to
be small, since $c<\sigma _{N}$.

The evolution time $t$ of the QRT algorithm is in the order of $1/cd_{1}$
such that the success probability of the algorithm is $O\left( 1\right) $.
It scales as $O\left( \kappa \right) $ since $d_{1}$ is in $O\left( 1\right)
$, and $\Delta ^{\ast }$ thus $c$ is in the order of $O\left( \kappa
^{-1}\right) $ in the worst case. The complexity of the QRT algorithm is
determined by Hamiltonian simulation of the algorithm $U(t)=\exp \left(
-iHt\right) $. By applying a $(\alpha ,m,0)$-block-encoding of $H$ in a
unitary matrix, the Hamiltonian simulation of $H$ is $U(t)=e^{-i\left(
H/\alpha \right) \alpha t}$, and $\alpha $ scales as $O\left( 1\right) $.
The optimal approach for Hamiltonian simulation is by applying the QSP
algorithm, for which the query complexity scales as $\Theta \left( \alpha t+%
\frac{\log \left( 1/\epsilon \right) }{\log \left( e+\frac{\log \left(
1/\epsilon \right) }{\alpha t}\right) }\right) $~by accessing the
Hamiltonian $H$~\cite{grandunif}. The number of times the Hamiltonian $H$ is
queried scales as $O\left[ \kappa +\log \left( 1/\epsilon \right) /\log \log
\left( 1/\epsilon \right) \right] $ since $t$ scales as $O\left( \kappa
\right) $ and $\alpha $ scales as $O\left( 1\right) $. The query complexity
of the algorithm by querying the oracles that access the Hamiltonian matrix $%
H$ scales as $O\left[ s\kappa +\log \left( 1/\epsilon \right) /\log \log
\left( 1/\epsilon \right) \right] $~\cite{QSP} for an $s$-sparse matrix $C$.
The total number of qubits required in this algorithm is $\left(
n+m+2\right) $.

\section{Discussion}

In this work, the problem of solving the linear system of equations is
transformed to finding the eigenstate of a Hermitian matrix with eigenvalue $%
0$. The QEF and QRT algorithms are used for obtaining this state by setting
an initial state whose overlap with the desired eigenstate is $O\left(
1\right) $. Because of this property, our algorithm have a better complexity
scaling compare with other algorithms where an initial state needs to be
prepared through an AQC procedure. The QEF algorithm is applied directly on
the initial state to project out the solution state without using the AQC or
QZE, thus achieves a better query complexity of $O\left[ s\kappa \log \left(
1/\epsilon \right) \right] $ than that of the algorithm in~\cite{Lin}. The
QRT algorithm evolves the initial state to the solution state of the QLSP
with query complexity of $O\left[ s\kappa +\log \left( 1/\epsilon \right)
/\log \log \left( 1/\epsilon \right) \right] $. The complexity of both
algorithms have linear scaling in $\kappa $, which is optimal since the
dependence on $\kappa $ cannot be made sublinear~\cite{HHL}. The complex
procedures such as phase estimation, amplitude amplification, AQC or QZE
used in previous algorithms are not needed, only time-independent
Hamiltonian simulation is used, for which the QSP algorithm provides an
optimal solution, thus our algorithms are easier to implement
experimentally. In the algorithms, we obtain the state $|\mathbf{v}%
_{N+1}\rangle $ that contains the solution vector of LSE as $\mathbf{v}%
_{N+1}=d_{0}\left(
\begin{array}{cc}
\mathbf{x^{\text{\textbf{T}}}/\left\Vert \mathbf{x}\right\Vert ,} & 0%
\end{array}%
\right) ^{\text{\textbf{T}}}+d_{1}\left(
\begin{array}{cc}
\mathbf{0^{\text{\textbf{T}}},} & 1%
\end{array}%
\right) ^{\text{\textbf{T}}}$. The contribution of the state $\left(
\begin{array}{cc}
\mathbf{0^{\text{\textbf{T}}},} & 1%
\end{array}%
\right) ^{\text{\textbf{T}}}$ in applications, such as calculating
expectation value of some operators, can be removed through some adjustment.

The linear solver is a basic engine in engineering and scientific computing,
and has wide applications in many areas. It paves a way for quantum machine
learning, and acts as an important ingredient in linear regression, Bayesian
inference, least-squares fitting, least squares support vector machine. The
numerical solvers for partial differential equations and ordinary
differential equations are also built on it. After numerical discretization,
such as the finite element method, finite difference method, or finite
volume method, one usually needs to solve a sparse linear system. The
quantum algorithms we presented here can be used as a subroutine in solving
these problems.

\begin{acknowledgements}
This work was supported by the National Key Research and Development Program of China~(No. 2021YFA1000600), the Natural Science Fundamental Research Program of Shaanxi Province of China under grants 2022JM-021 and the Fundamental Research Funds for the Central Universities~(Grant No.~11913291000022).
\end{acknowledgements}

\end{document}